\documentclass[prl,aps,showpacs,twocolumn]{revtex4}

\def\be{\begin{equation}}
\def\ee{\end{equation}}

\usepackage{amsmath}
\usepackage{amsfonts}
\usepackage{graphicx}
\usepackage{bm}
\usepackage[latin1]{inputenc}

\begin{document}
\title{
Condensation Energy of a Spin-1/2 Strongly Interacting Fermi Gas}
\author{N. Navon$^1$\footnote{Present email and address: nn270@cam.ac.uk, Cavendish Laboratory, University of Cambridge, J.J. Thomson Ave., Cambridge CB3 0HE, United Kingdom}, S. Nascimb\`ene$^1$, X. Leyronas$^2$\footnote{leyronas@lps.ens.fr}, F. Chevy$^1$ and C. Salomon$^1$}
\affiliation{$^1$Laboratoire Kastler Brossel, CNRS, UPMC, Ecole Normale Sup\'erieure, 24 rue Lhomond, 75231 Paris, France\\
$^2$Laboratoire de Physique Statistique, Ecole Normale Sup\'erieure, UPMC
Univ Paris 06, Universit\'e Paris Diderot, CNRS, 24 rue Lhomond, 75005 Paris,
France.}
\date{\today}

\begin{abstract}
We report a measurement of the condensation energy of a two-component Fermi gas with tunable interactions. From the equation of state of the gas, we infer the properties of the normal phase in the zero-temperature limit. By comparing the pressure of the normal phase at $T=0$ to that of the low-temperature superfluid phase, we deduce the \emph{condensation energy}, i.e. the energy gain of the system in being in the superfluid rather than normal state. We compare our measurements to a ladder approximation description of the normal phase, and to a fixed node Monte-Carlo approach, finding excellent agreement. We discuss the relationship between condensation energy and pairing gap in the BEC-BCS crossover.
\end{abstract}

\maketitle

From a thermodynamic point of view, a superconducting state is favored compared to a normal state when the free energy of the former ($E_S$) is lower than the latter ($E_N$). This energy difference, called \emph{condensation energy} is a central concept in the BCS theory of conventional superconductivity. For example, in the weakly interacting regime the condensation energy is related to the superfluid pairing gap $\Delta$ by
\be\label{EBCSDelta}
E_c=E_S-E_N=-N_f\frac{\Delta^2}{2}
\ee
where $N_f$ is the density of states at the Fermi energy \cite{schrieffer1999theory}. For superconductors, the condensation energy is obtained from the measurement of the critical magnetic field $H_c$ at which superconductivity is quenched
\be
E_c=-\mu_0\frac{H_{c}^2}{2}\label{eqHc}
\ee
where $\mu_0$ is the vacuum magnetic permeability \cite{schrieffer1999theory}.
 While BCS theory (and relation (\ref{EBCSDelta})) have proven very successful to explain conventional superconductivity, a similar description to explain exotic forms of superconductivity, such as encountered in cuprate or iron-compound materials is still lacking. In particular, the role of the condensation energy in high-$T_c$ superconductors is thought to give insight on the mechanism that could be responsible for driving the superconducting transition (see for example \cite{leggett1996interlayer,demler1998quantitative,scalapino1998superconducting,haslinger2003condensation} and references therein), though the method to extract it from experimental data is still a hotly debated issue \cite{chakravarty2003condensation,van2002condensation,chakravarty1999frustrated}.

Ultracold atoms are now increasingly used as testbeds to experimentally explore quantum many-body physics, owing to their high degree of control \cite{bloch2008many}. It now becomes possible to simulate hamiltonians from various fields of physics, such as neutron matter or condensed matter physics in simple systems. Moreover, interactions between ultracold atoms, characterized by the $s$-wave scattering length $a$, can be tuned via magnetic Feshbach resonances, giving access to the regime of strong interactions.

In this letter, we investigate the condensation energy of a dilute spin-1/2 strongly interacting  Fermi gas with variable interaction strength.
We show that the condensation energy can be measured by applying a chemical potential imbalance between the two spin states which is the analogue
of a magnetic field in superconductors. We compare our experimental results to a diagrammatic theory, finding excellent agreement.

The experimental setup was presented in \cite{nascimb2010exploring}. Our system is a quantum gas of $^6$Li prepared in a mixture of its two lowest energy spin states.
  The gas is loaded in a single beam dipole trap, providing a radial (strong) confinement, while the axial (weak) confinement ($z$-axis) is provided by magnetic coils. This results in a cigar-shaped trap. The interactions are tuned using a pair of coils in Helmholtz configuration in order to create the large homogeneous bias field to tune the scattering length $a$ via the 832.18 G Feshbach resonance \cite{jochim2012Feshbach}. The mixture is cooled to quantum degeneracy by lowering the trap depth and absorption images perpendicular to the weak direction are recorded to obtain the in-situ density distributions along the $z$-axis. Previous theoretical \cite{cheng2007trapped,ho2009obtaining} and experimental studies \cite{nascimb2010exploring,navon2010equation} have demonstrated that the density profiles of a trapped spin-imbalanced Fermi gas can be used to extract the Equation of State (EoS) of the corresponding homogeneous system via the pressure formula: $P(\mu_1,\mu_2,T)=\frac{m\omega_r^2}{2\pi}(\bar{n}_1(z)+\bar{n}_2(z))$, where $\omega_r$ is the radial trapping frequency, and $\bar{n}_i(z)=\int\textrm{d}^2r\;n_i(r,z)$ is the doubly-integrated density distribution of spin-species $i$ ($i=1,2$).
\begin{figure}[h!]
\begin{center}
\includegraphics[width=0.9\columnwidth]{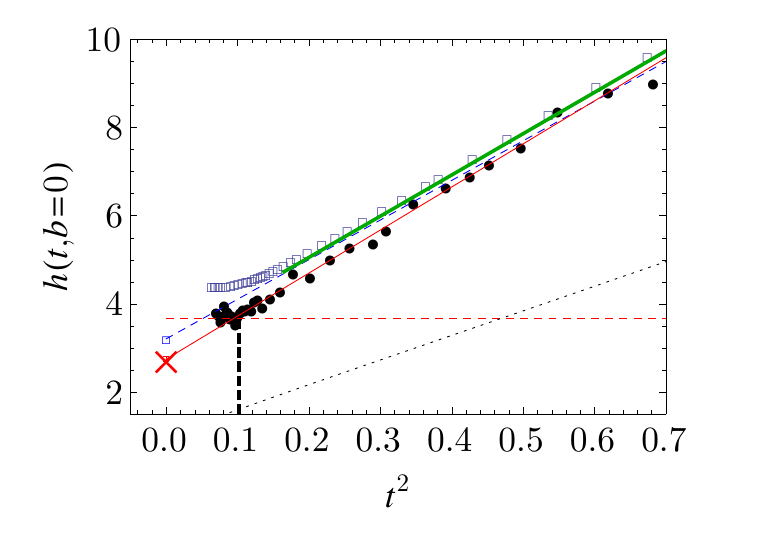}
\end{center}
\caption{
(Color online) Reduced pressure $h(t,b)=P(\mu_1,\mu_2,T)/2P_0(\bar{\mu})$ of the spin-1/2 unitary Fermi gas, where $P_0$ is the $T=0$ Fermi pressure of an ideal gas, $t$ the reduced temperature $k_BT/\mu$, and $b=0$ ie unpolarized gas.  Black points are from from \cite{nascimb2010exploring}. The Fermi liquid fit is shown as a continuous red line, and the extrapolated zero temperature pressure of the normal state $\xi^{-3/2}_N$ is signaled by a red cross. Data from \cite{ku2012revealing} is shown in empty blue squares, the corresponding fit in dashed blue line, and the extrapolation at $t=0$ in empty blue square. Bold diagrammatic Monte Carlo calculation \cite{van2012feynman} is shown in solid green line. the vertical dashed line signals the normal/superfluid transition at $t_c=0.33$.
} \label{Fig1}
\end{figure}
\begin{figure}[h]
\begin{center}
\includegraphics[width=.8\columnwidth]{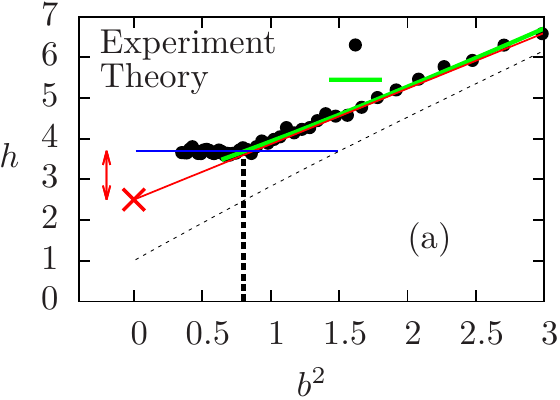} \\
\includegraphics[width=0.85\columnwidth]{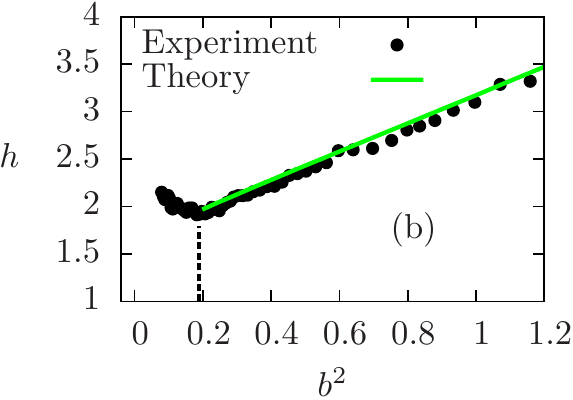} \\
\includegraphics[width=0.8\columnwidth]{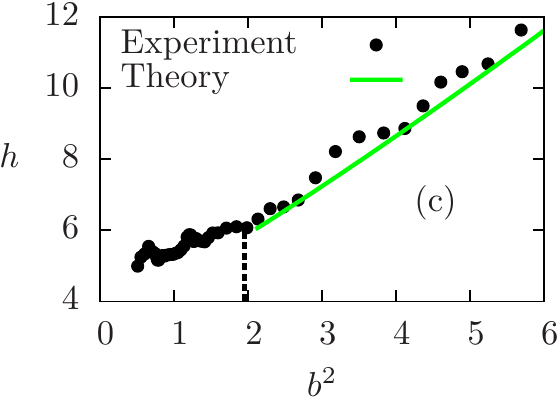}
\end{center}
\caption{Pressure of the spin-imbalanced gas in the BEC-BCS crossover at $t=0$. The position of the first-order phase transition to the superfluid is shown by a  vertical dashed black line. (a) Unitary limit. The Fermi liquid fit is shown as a solid red line, while the $t=0$ Equation of State in the superfluid phase is in blue solid line. The pressure of the non-interacting gas is displayed as a dotted black line. The $t=0$ and $b=0$ extrapolation of the normal phase pressure is shown as a red cross, and the condensation pressure as the double-arrows. (a), (b), and (c): results of the ladder approximation for the normal phase are shown in green for $\delta=0,\,-0.58$ and $+0.2$ respectively.}
\label{fighbdelta}
\end{figure}

At unitarity, where the scattering length $a$ diverges, we previously measured the pressure of the spin-balanced gas as a function of the reduced temperature $t=k_B T/\mu$ (where $2\mu=\mu_1+\mu_2$) \cite{nascimb2010exploring}, as well as the pressure of the spin-imbalanced gas at $t\approx0$ as a function of the spin-polarizing field $b=\frac{\mu_1-\mu_2}{\mu_1+\mu_2}$. We suggested that the low-temperature properties of the normal phase of the Fermi gas were consistent with a Fermi-liquid behavior \cite{nascimbene2011fermi}. As a result, the low-temperature and low-imbalance limit of the pressure of the unitary gas can be written as:
\be\label{FermiLiqEq}
h(t,b)=\frac{P(\mu_1,\mu_2,T)}{2P_0(\mu)}\simeq\xi_N^{-3/2}+\frac{\tilde{\chi}b^2}{2}+\frac{\tilde{c}_V t^2}{2}
\ee
where $P_0(\mu)=\frac{1}{15\pi^2}\left(\frac{2m}{\hbar^2}\right)^{3/2}\mu^{5/2}$ is the ideal Fermi gas pressure. The response coefficient to temperature $t$ is the dimensionless specific heat $\tilde{c}_V$, while the response to the polarizing field $b$ is the dimensionless magnetic susceptibility $\tilde{\chi}$ (respectively equal to $5\pi^2/8$ and $15/4$ for the ideal Fermi gas). The magnetic susceptibility has been the subject of a previous work \cite{nascimbene2011fermi}, and we focus here on the measurement of pressure of the normal phase $\xi_N^{-3/2}$ in the $t=0$ and $b=0$ limits. In the ($t,b$) plane, our measurements of the EoS of the unitary gas have been performed along two directions: the unpolarized gas as a function of temperature $h(t,b=0)$ (Fig.\ref{Fig1}), and the low-temperature polarized gas versus chemical potential imbalance $h(t=0,b)$ (Fig.\ref{fighbdelta}(a)). The quadratic behavior of the pressure versus both $b$ and $t$ supports the Fermi-liquid interpretation of the low-temperature thermodynamic properties of the normal phase. However, the system will ultimately undergo a second order phase transition to a superfluid state and, below a temperature $t_c\sim 0.33$, the pressure of the spin-balanced gas deviates from the $t^2$ behavior. In contrast at $t=0$, the spin-imbalanced gas ($\mu_1 \neq \mu_2$) undergoes a first-order phase transition to an unpolarized superfluid phase when
$h_S(0,0)=h_N(0,0)+\tilde{\chi}b^2/2$. This condition is the analogue of Eq.\ref{eqHc}, and at unitarity it yields the critical chemical potential imbalance $b_c\approx\sqrt{0.8}$, see Fig.\ref{fighbdelta}(a). This is testified by the discontinuity in the slope of $h$ vs $b^2$. From Eq.(\ref{FermiLiqEq}), and extrapolating the Fermi-liquid behavior to the zero-temperature and spin-balanced limits, we measure the $T=0$ dimensionless pressure of the spin-balanced unitary gas in the normal phase $\xi_N^{-3/2}$. In the first limit $(t\rightarrow0,b=0)$ we find $\xi_N=0.51(2)$, while in the second one $(t=0,b\rightarrow0)$, we extract $\xi_N=0.53(2)$ (see the red crosses in Fig.\ref{Fig1} and Fig.\ref{fighbdelta}(a)). The agreement between these values, taken for two very different limiting regimes, is remarkable and further supports an accurate description of the normal phase of the unitary gas as a Fermi liquid. This value is in good agreement with values calculated using Monte Carlo methods 0.54 \cite{carlson2003superfluid}, 0.56 \cite{lobo2006nsp}, and 0.52 \cite{bulgac2008quantum}. It is somewhat larger than a recent experimental determination, 0.46(1) \cite{ku2012revealing}.

The problem of the zero temperature balanced superfluid Fermi gas has been the subject of thorough theoretical investigations \cite{zwergerbook}. However, much less work has been
devoted to the equation of state of the zero temperature normal phase \cite{lobo2006nsp}.
We show below that our experimental results can be quantitatively reproduced using the ladder approximation \cite{PS,clk}. This theory includes the repeated two-body scattering between particles $1$ and $2$ described by the scattering length $a$. In particular, for $a^{-1}>0$, it contains the physics of a molecular state. We use the finite temperature formalism and take the zero temperature limit.
The self-energy for the particles $2$, which physically describes the effect of interaction between particles is given by (we take $\hbar=1$)
\begin{eqnarray}
\Sigma_2(k,i\omega)&=&\int\!\!\frac{d^3{\bf K}}{(2\pi)^3}\!\int_{i\mathbb{R}}\!\frac{d\Omega}{2\pi i}
\frac{
\Gamma(K,\Omega)
}
{
\left[
\Omega-i\omega+\mu_{1}-\frac{({\bf K}-{\bf k})^2}{2 m}
\right]
}\label{eqSigma2}
\end{eqnarray}
where the two-particle vertex $\Gamma$ is given by
\begin{eqnarray}
\Gamma(K,\Omega)^{-1}&=&\frac{m}{4\pi\,a}+\Pi(K,\Omega)
\end{eqnarray}
 where $\Pi(K,\Omega)$ is the pair bubble \cite{clk}. At zero temperature,  $\Pi(K,\Omega)$ can be calculated analytically.
 The pairing instability, signaling a second order phase transition, is found using the Thouless criterion $\Gamma^{-1}(0,0)=0$. For given $\mu_1$ and $a$, this happens for a critical
 value of the chemical potential  $\mu_{2c}$ of particles $2$. In order to stay in the normal phase, we have performed our calculations for $\mu_2<\mu_{2c}$.
The integration on $\Omega$ can be performed by deforming the integration contour in the half-plane $\Re(\Omega)<0$. In this way, we pick the singularities of the integrand in Eq.\ref{eqSigma2} and get three contributions corresponding to the pole of $(\Omega-i\omega+\mu_{1}-\frac{({\bf K}-{\bf k})^2}{2 m})^{-1}$ ($\Sigma_L$), the branch cut of $\Gamma(K,\Omega)$
($\Sigma_{\Gamma}$) and the molecular pole $\Omega_0(K)$ (for $a^{-1}>0$) of $\Gamma(K,\Omega)$ ($\Sigma_m$) \cite{clk}.
$\Omega_0(K)+2\mu$ represents physically the energy of a molecule of momentum ${\bf K}$. We find that in the normal phase
$\Omega_0(K)>0$.
As a consequence, when we deform the integration contour in $\Re(\Omega)<0$, we do not get any contribution from the molecular pole of $\Gamma$, and therefore we have $\Sigma_m=0$.
This is consistent with the physical argument in favor of the absence of molecule in the normal phase. Indeed, if we had some molecules in the system, they would be
condensed at zero temperature. Therefore the system would be superfluid, and we are no longer entitled to use Eq.\ref{eqSigma2}.
We deduce the minority density $n_2$ using the Fermi liquid type relation due to Landau
\begin{eqnarray}
\mu_{2}&=&\frac{k_{F,2}^2}{2m}+\Sigma_{2}(k_{F,2},0)\label{eqLandau}
\end{eqnarray}
where by definition $k_{F,2}\equiv (6\pi^2\,n_{2})^{1/3}$, is the Fermi wave vector of particles of type $2$. For given $\mu_1$, $\mu_2$ and $a$, this is an implicit equation for $k_{F,2}$, hence $n_2$.
As it was found in Ref.\cite{crlc}, we find a non zero density $n_2$ for a chemical potential $\mu_2$ larger than the {\it polaron} \cite{crlc}\cite{polandco} chemical potential
$\mu_{p}(\mu_1)$.
In practice, we fix $\mu_1>0$, then we solve Eq.\ref{eqLandau} for a given $\mu_2\geq \mu_{p}(\mu_1)$. The pressure is determined by integrating the density using the Gibbs-Duhem relation
\begin{eqnarray}
P(\mu_1,\mu_2)&=P_0(\mu_1)+\int_{\mu_p}^{\mu_2}d\mu'_2\frac{1}{6\pi^2}\left[k_{F,2}(\mu_1,\mu'_2)
\right]^3
\end{eqnarray}
For a fixed $\mu_1$, we calculate the minority density for increasing minority chemical potential between $\mu_p(\mu_1)$ and $\mu_2$. For a sufficiently large chemical potential difference, the system is normal (the pairing susceptibility does not diverge).
For sufficiently low $b$, we calculate the dimensionless equation of state $h(\delta,b)$, where $\delta$ is the grand-canonical interaction strength, $\delta=\hbar/\sqrt{2m\mu}a$. For all values of $\delta\leq0$, we find a linear behavior of $h$ as a function of $b^2$. The comparison between experiment and theory is shown for $\delta=0$ (Unitary limit), $\delta=-0.58$ (BCS side of the crossover) and $\delta=0.2$ (BEC side) in Figs.\ref{fighbdelta}(a),\ref{fighbdelta}(b) and \ref{fighbdelta}(c) respectively. The agreement is very good. 
However, for increasing $a^{-1}>0$, the values of $b$ in the normal phase become larger and larger, and as a consequence the linear fit of $h$ as a function of $b^2$, valid at low $b$, is worse. Still, for $\delta=0.2$ the experimental equation of state $h(\delta,b)$  is in good agreement with the ladder approximation calculation above $b_c$, curved green line in Fig.\ref{fighbdelta}(c).
 Within the ladder approximation we have determined the critical spin polarizing field $b_c$ at which a pole appears in the vertex function $\Gamma$ at zero frequency and zero wave vector   (Thouless criterion). We found that $b_c$ was always {\it smaller} than the experimental value of the first order transition. Our calculation is therefore free of any instability singularity in the normal phase.
 For the spin-susceptibility, we also find a good agreement between the ladder approximation, experiments, and Monte Carlo simulations of \cite{nascimbene2011fermi}.

\begin{figure}[ht!]
\centerline{\includegraphics[width=1.\columnwidth]{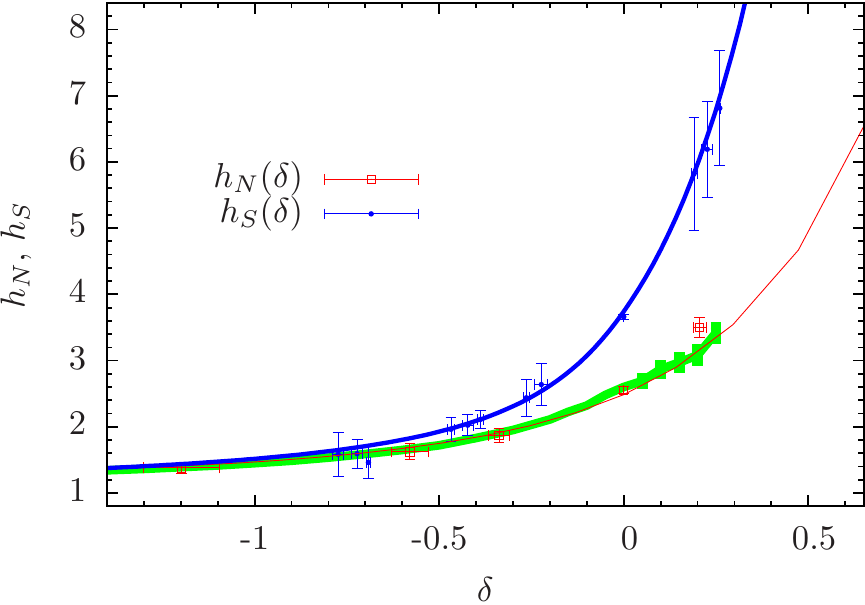}}
\caption{(Color online) Pressure of the normal $h_N$, red empty squares, and superfluid $h_S$, blue circles, phases at low temperature in the BEC-BCS crossover measured in \cite{navon2010equation}. The green line is the result of the ladder approximation. The solid blue line is a guide for the eye, while the red solid line is the result of Fixed-Node Monte Carlo calculations \cite{nascimbene2011fermi}. The difference between the blue and red/green lines is the {\it condensation pressure.}}\label{Fig3}
\end{figure}

\begin{figure}[h!]
\centerline{\includegraphics[width=0.9\columnwidth]{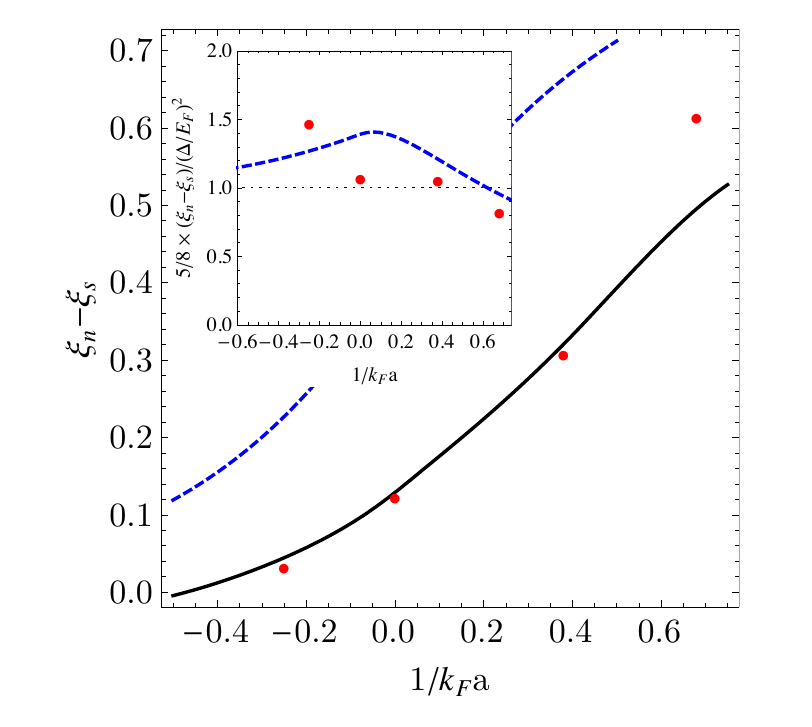}}
\caption{(Color online) Relation between the condensation energy and the superfluid pairing gap. Main panel: measured dimensionless condensation energy $\xi_N-\xi_S$ versus interaction strength $1/k_Fa$ (solid black line). In red points,  BCS expression (\ref{EcBCS}) using the values of $\Delta$ measured in \cite{schirotzek2008dsg}.  For comparison the prediction from mean-field BCS theory is shown in dashed blue line. A Fixed node Monte Carlo calculation \cite{nascimbene2011fermi} coincides with the solid black line. Inset: ratio of the condensation energy $\xi_N-\xi_S$ to $\frac{5}{8}\left(\frac{\Delta}{E_F}\right)^2$.} \label{Fig4}
\end{figure}

Gathering the results from Fig.2, we now extract the zero-temperature dimensionless pressure $h_N$ of the normal phase as a function of $\delta$ \cite{nascimbene2011fermi}. The resulting EoS of the normal phase $h_N(\delta)$ is plotted in Fig.\ref{Fig3} as red empty squares together with the ladder approximation calculation (green line), showing excellent agreement in the explored crossover. For comparison, the previously measured equation of state of the low-temperature gas in the superfluid phase $h_S(\delta)$ is shown as blue points and a blue line fit \cite{navon2010equation}. The difference between the superfluid and normal pressure at $T=0$ thus represents the {\it condensation pressure}. The superfluid pressure is higher than the normal phase pressure, $h_S(\delta)>h_N(\delta)$, hence the grand-potential is lower and the superfluid state is the stable phase at low-temperature. Turning to the canonical ensemble the superfluid and normal phase energies $\xi_S$ and $\xi_N$ as a function of the canonical interaction strength $1/k_Fa$ can be computed from the pressure measurement of Fig.\ref{Fig3} using a Legendre transform \cite{supplementScience}. The measured condensation energy $\xi_N-\xi_S$ is shown as a solid black line in Fig.\ref{Fig4}. 

In the BCS regime, the condensation energy $E_c$ can be  explicitly calculated from the energy of the superconducting and normal states,  yielding the well known result $E_c=\frac{3}{8}N\frac{\Delta^2}{E_F}$,
where $\Delta$ is the single-particle excitation gap, and $E_F$ the Fermi energy. Since $E=\frac{3}{5}NE_F\xi_\alpha(1/k_Fa)$ (where $\alpha=S,N$), the BCS equation becomes:
\be\label{EcBCS}
\xi_N-\xi_S=\frac{5}{8}\left(\frac{\Delta}{E_F}\right)^2.
\ee

Strictly speaking, this formula is valid only in the weakly attractive limit $\Delta\rightarrow 0$. For an arbitrary interaction, the condensation energy is given by a more involved function of the gap and, based on dimensional arguments, it  should be written as
\be
\xi_N-\xi_S=\frac{5}{8}\left(\frac{\Delta}{E_F}\right)^2F(\Delta/E_F),
\ee
where $F$ is a (yet) unknown function with $F(0)=1$ to satisfy the BCS prediction. In the spirit of Landau's theory, the $U(1)$ invariance  suggests that $F$ can be expanded with $(\Delta/E_F)^2$ and as such, the first beyond-BCS correction should be proportional to $|\Delta/E_F|^2$. At unitarity where $\Delta \simeq 0.5 E_F$ \cite{carlson2003superfluid} this leads to a moderate 25\% correction to the BCS prediction, which suggests that the range of validity of Eq. \ref{EcBCS} should extend beyond the strict weakly interacting limit \cite{bulgac2008quantum}.

In order to test the BCS expression (\ref{EcBCS}) in the BEC-BCS crossover, we compare our measurement of the condensation energy  to  $\frac{5}{8}\left(\frac{\Delta}{E_F}\right)^2$  using the values of $\Delta$ measured by radio-frequency spectroscopy in \cite{schirotzek2008dsg}, red points in Fig.\ref{Fig4}. The agreement seen in Fig.\ref{Fig4} indicates that, even in the strongly interacting regime, the BCS expression is remarkably valid. 
A more stringent test is provided by plotting the ratio between left-hand and right-hand side of Eq.(\ref{EcBCS}) (inset of Fig.\ref{Fig4}) and
 we indeed find a ratio close to unity. Note that calculating this ratio using BCS mean-field theory provides a reasonable estimate (dashed blue line in inset of Fig. \ref{Fig4}) even though the absolute values of the condensation energy (dashed blue line in the main panel of Fig.\ref{Fig4}), or of the pairing gap are both quantitatively inaccurate in the strongly interacting regime.

In summary, we have measured the condensation energy of a two-component Fermi gas with tunable interactions.
The temperature and spin-polarizing field dependence of the normal phase pressure are in good agreement  with a Fermi liquid description.
A simple ladder approximation calculation quantitatively reproduces experimental data at zero temperature in the normal phase. Future work will explore the critical region and search for exotic phases such as the FFLO phase \cite{zwergerbook}.


We thank M. Ku, and M. Zwierlein for providing their experimental data, and F. Werner for the bold diagrammatic Monte Carlo calculations. We acknowledge support from Institut de France (Louis D.), R\'egion Ile de France (IFRAF-C'nano), ERC Ferlodim, and Institut Universitaire de France. 



\end{document}